\documentclass[12pt]{article}

\usepackage[utf8]{inputenc}
\usepackage[T1]{fontenc}

\usepackage{setspace} 
\usepackage[vmargin = 1.5in, hmargin = 1.5in]{geometry}

\usepackage[usenames,dvipsnames,svgnames]{xcolor}
\usepackage{amsmath, amssymb, amsfonts, graphicx, tikz,pdflscape, mathtools, amsthm, upgreek, bm,pgfplots,csvsimple,multirow, multicol, booktabs}
\usepackage{verbatim}
\usepackage{natbib}
\usepackage{accents}

\usepackage{needspace}

\usepackage{comment}
\usepackage[inline]{enumitem}
\usepackage{tocvsec2}
\usepackage{threeparttable}
\usetikzlibrary{calc, cd}
\setlist{noitemsep, topsep=4pt, listparindent = \parindent}
\usepackage[skip = 10pt]{caption}
\allowdisplaybreaks[1]
\allowdisplaybreaks[1]

\definecolor{Blue}{RGB}{86,180,233}
\definecolor{Orange}{RGB}{230,159,0}
\definecolor{Green}{RGB}{0,158,115}
\definecolor{GmailBlue}{RGB}{42, 93, 176} 
\usepackage[
pagebackref,
colorlinks=true,
citecolor= GmailBlue,
linkcolor=GmailBlue,
urlcolor = GmailBlue
]{hyperref}

\newcommand{\bibtexorder}[1]{}

\usepackage{pgfplots}
\pgfplotsset{compat=newest}
\usepgfplotslibrary{groupplots}
\usepackage{tikz}
\usetikzlibrary{matrix,calc,shapes,arrows.meta,positioning}
\pgfplotsset{width = \textwidth/2}
\tikzstyle{hollow}=[circle,draw,inner sep=1.5]
\tikzstyle{solid}=[circle,draw,inner sep=1.5,fill=black]
\pgfplotsset{compat = newest}

\usepackage[capitalize,noabbrev]{cleveref}

\newtheoremstyle{breakital}
{}
{}
{\itshape}
{}
{\bfseries}
{}
{\newline}
{}

\theoremstyle{breakital}
\newtheorem{thm}{Theorem}
\newtheorem*{theorem*}{Theorem}
\newtheorem*{cor*}{Corollary}

\newtheorem{lem}{Lemma}

\crefname{prop}{Proposition}{Propositions}
\crefname{thm}{Theorem}{Theorems}
\crefname{lem}{Lemma}{Lemmas}

\newtheoremstyle{break}
{}
{}
{}
{}
{\bfseries}
{}
{\newline}
{}

\theoremstyle{break}

\crefname{as}{Assumption}{Assumptions}

\theoremstyle{definition}

\newtheorem*{rem*}{Remark}

\numberwithin{lem2}{section}
\crefname{lem2}{Lemma}{Lemmas}

\def\l{\lambda}

\def\R{\mathbf{R}}

\def\F{\mathbf{F}}

\newcommand{\Paren}[1]{\left( #1 \right)}

\usepackage{datetime}
\newdateformat{specialdate}{\THEDAY~\monthname[\THEMONTH] \THEYEAR}

\title{A Unified Theorem of the Alternative for Linear Inequalities}
\author{Ian Ball\thanks{I thank John Geanakoplos for first teaching me about Farkas' Lemma. For comments, I thank Stephen Morris, Joseph Root, Iv\'{a}n Werning, and the audience at MIT IAP 2023.}}
\date{\specialdate\today}

\begin{document}
	
\maketitle

\begin{abstract}
	This note presents a unified theorem of the alternative that explicitly allows for any combination of equality, componentwise inequality, weak dominance, strict dominance, and nonnegativity relations. The theorem nests 60 special cases, some of which have been stated as separate theorems. 
\end{abstract}

Many results in economic theory can be proven by applying the separating hyperplane theorem to two carefully chosen convex sets: the fundamental theorem of asset pricing \citep[e.g.,][p.~4]{Duffie}; the never-best-response/strict-dominance equivalence in noncooperative game theory \citep{Pearce, Weinstein};   the Bondareva--Shapley theorem on the nonemptiness of the core in balanced coalitional games  \citep[e.g.,][pp.~262--263]{OsborneRubinstein}; full-surplus extraction in auctions with correlated types \citep{CM}; and Border's characterization of feasible reduced-form auctions \citep{Border2007}. Whenever the proof involves finite systems of  \emph{linear} equalities and inequalities, as in these examples, a more direct proof is available using the \emph{theorem of the alternative}---a purely algebraic result that does not rely on calculus or limits. The theorem of the alternative has many different versions, associated with Farkas, Freedholm, Gale, and many others.\footnote{Two closely related mathematical results are the minimax theorem and linear programming duality. For economic applications beyond  zero-sum games and linear programming, the theorem of the alternative is usually more convenient to use than these other two results.  It has long been known that the theorem of the alternative and linear programming duality can each be used to directly prove the other two results. Starting from the minimax theorem is more subtle. \cite{RenyBrooks} complete the equivalence by proving the linear programming duality directly from the minimax theorem.} It is hard to keep track of them all, making it inconvenient to find the right form for a particular application.

I present a single, unified theorem of the alternative that explicitly allows for any combination of equality ($=$), componentwise inequality ($\leq$), weak dominance ($<$), strict dominance ($\ll$), and nonnegativity relations.\footnote{That is, $x \leq y$ means $x_i \leq y_i$ for all $i$; $x < y$ means $x \leq y$ and $x \neq y$; and $x \ll y$ means $x_i < y_i$ for all $i$.}  Putting these different relations together reveals an intuitive structure between the primal system and its dual. I hope that this single, unified version is easier to remember (or at least to apply) than the myriad versions in the literature. 

\begin{thm}[Unified theorem of the alternative] \label{res:unified}
	For $i \in \{ 1, 2, 3, 4\}$, fix a matrix $A_i \in \R^{m_i \times n}$ and a vector $b_i \in \R^{m_i}$. Exactly one of the following holds. 
		\begin{enumerate}
			\item[P.] There exists $x \in \R^n$ satisfying
			\begin{align}
				A_1 x &= b_1, \\
				A_2 x &\leq b_2, \\
				A_3  x  &< b_3, \\
				A_4 x &\ll b_4, \\
				(x &\geq 0). \label{it:N} \tag{N}
			\end{align}
			\item[D.] There exist $\l_i \in \R^{m_i}$ for $i = 1,2,3,4$ satisfying\footnote{A comma between statements in a system means \emph{and}. The word \emph{or} is written out.}
			\begin{align}
				&\l_i \geq 0~\text{for}~i = 2,3,4, \label{it:M} \tag{M} \\
				&\sum_{i=1}^{4} \l_i^T A_i \mathrel{ \mathop{=}\, (\geq)} 0, \label{it:L} \tag{L}\\
				&\sum_{i=1}^{4} \l_i^T b_i \leq 0, \label{it:R} \tag{R}\\
				&\sum_{i=1}^{4} \l_i^T b_i < 0 ~\text{or}~ \l_3 \gg 0 ~\text{or}~ \l_4  > 0. \label{it:S} \tag{S}
			\end{align}
		\end{enumerate}
	\noindent If the primal system includes relation $i$ only for $i$ in a subset $I$ of $\{1, 2,3,4\}$, then in the dual system, set $\l_j = 0$ for $j \notin I$, leaving unknowns $\l_i$ for $i \in I$. 
\end{thm}

The expressions in parentheses mean that if the nonnegativity constraint \eqref{it:N} is included in the primal system, then \eqref{it:L} in the dual system becomes an inequality, rather than an equality. 

Any solution of the dual system certifies the infeasibility of the primal system. To see this, for each $i = 1,2,3,4$, left multiply relation $i$ by some vector $\l_i^T$, with $\l_i \geq 0$ if $i \geq 2$. Sum to get 
\begin{equation} \label{eq:contradiction}
	\Paren{\sum_{i=1}^{4} \l_i^T A_i } x \leq \sum_{i=1}^{4} \l_i^T b_i, 
\end{equation}
with strict inequality if $\l_3 \gg 0$ or $\l_4 > 0$. To have any chance at a contradiction, the vector coefficient on the left side must be zero, and the scalar on the right side must be weakly less than $0$. Then we get a contradiction if the right side is strictly negative or if the inequality is strict, i.e., if $\l_3 \gg 0$ or $\l_4 > 0$. These conditions constitute the dual system, which includes constraints on the multipliers \eqref{it:M} and  on the left and right sides of \eqref{eq:contradiction}, denoted \eqref{it:L} and \eqref{it:R}, together with a strictness condition \eqref{it:S}. 

The harder direction of \cref{res:unified} says that if the primal system is infeasible, then there exists a certificate of its infeasibility. The proof (\cref{sec:proof_unified,sec:proof_basic}) is purely algebraic.  \cref{res:unified} is derived from a more basic theorem of the alternative, which is proven by Fourier--Motzkin elimination.  This proof shows that \cref{res:unified} holds with any ordered field in place of the set $\R$.\footnote{\cite{Border} states separate versions of the theorem of the alternative with rational coefficients and dual variables.}

The primal and dual inequalities have an intuitive structure---more restrictive primal relations are easier to contradict. There is a sign restriction on $\l_2$ (associated with the componentwise inequality) but not on $\l_1$ (associated with the more restrictive equality relation). As $i$ increases from $2$ to $3$ to $4$, the stricter the primal inequality, the more permissive is the constraint on the associated dual variable. 

By dropping some relations from the primal system, we can immediately read off from \cref{res:unified} all of the standard theorems of the alternative. If the primal system contains relation $i$ only for $i$ in some subset $I$ of $\{1, 2, 3, 4\}$, then setting $\l_j = 0$ for all $j \not\in I$ in the dual system amounts to (a) restricting each summation to $i$ in $I$, and (b) removing any other inequality that contains a single multiplier $\l_j$ with $j \notin I$.\footnote{We can remove true statements from conjuctions  and false statements from disjunctions. Even if $I = \varnothing$, we get the primal  $(x \geq 0)$ and the dual $0 < 0$, so the statement remains correct.} For example, setting $I = \{2\}$ and dropping subscripts, the primal system (without nonnegativity) becomes
\[
A x \leq b,
\]
and the dual system becomes
\[
\l \geq 0, \qquad \l^T A = 0,  \qquad \l^T b < 0.
\]
This basic form of the theorem of the alternative can be found in \cite{Gale}. \cref{res:unified} immediately yields  $60 = 15 \cdot 2 \cdot 2$ distinct theorems of the alternative, corresponding to the $15$ nonempty collections of the four relations $=, \leq , < , \ll$, with or without the nonnegativity constraint, in homogeneous ($b = 0$) and inhomogeneous form.

\cref{table:alt} lists the theorems of the alternative appearing in two popular references: \cite{Border} and \cite{Perng}.\footnote{I also include a result appearing in John Geanakoplos's lecture notes, which he calls ``Generalized Farkas' Lemma.'' The theorems are organized according to the relations that are allowed in the primal system. I do not distinguish the direction of the inequalities (e.g., $\leq$ versus $\geq$) since the direction can  be flipped by negating $A_i$ and $b_i$.} Each theorem in \cref{table:alt} is one of the 60 special cases of \cref{res:unified}, with the exception of Morris, which requires additional manipulations.\footnote{In \cite{Morris}, the primal system is a disjunction over a family of systems of linear inequalities. He proves the result by applying a different theorem of the alternative (Antosiewicz II in \cref{table:alt}). Morris calls that result Farkas's Lemma and cites \cite{Gale} for the proof.}  In these results, it is not immediately clear how to modify the dual when some inequalities are dropped from the primal.\footnote{Indeed, the statements of the theorems of Tucker, Motzkin, and Slater in \cite{Perng}, following \cite{Mangasarian}, explicitly require that the matrices appearing in the strict inequalities ($<$ and $\ll$) are ``nonvacuous.'' \citet[p.~28]{Mangasarian} claims that this restriction is essential for his proof. In fact, the results and proof and go through without these restrictions.} In contrast, \cref{res:unified} includes a simple procedure for dropping inequalities from the primal. Therefore, there is no need to keep track of the special cases. 

\begin{table}
	\begin{center}
		\begin{tabular}{l ccc } 
			Author &  ~$=, \leq, <, \ll$~ & ~$x \geq 0$~  & ~$b = 0$~ \\
			\toprule
			Freedholm &$=$ & N& N \\
			Farkas & $=$ & Y & N \\
			\midrule
			Gale  I & $\leq$ & N & N \\
			Gale  II& $\leq$ & Y & N \\
			\midrule
			Stiemke & $<$ & N & Y \\
			Mangasarian & $<$ & Y & N \\
			\midrule
			Gordan  & $\ll$ & N & Y \\
			Ville & $ \ll$ & Y & Y \\
			\midrule
			Farkas II & $=$, $\leq$ & Y& N \\
			Antosiewicz I & $\leq, <$ & N & Y \\
			Antosiewicz II & $\leq, \ll$ & N & Y \\
			\midrule
			Morris & $\leq, \ll$ & N & Y \\
			\midrule
			Tucker & $=, \leq, <$ & N & Y \\
			Motzkin & $=, \leq, \ll$ & N  & Y \\
			Slater  & $=, \leq, <, \ll$ &  N & Y \\
			Geanakoplos & $=, \leq, \ll$ & N& N
		\end{tabular}
	\end{center}
	\caption{Theorems of the alternative}
	\label{table:alt}
\end{table}

appendix

\section{Proof of unified version from basic version} \label{sec:proof_unified}

Here I derive \cref{res:unified} from the following standard theorem of the alternative (\cref{res:basic}) which appears in \cite{Gale}. For completeness, I include a proof of \cref{res:basic} using Fourier--Motzkin elimination in \cref{sec:proof_basic}.

\begin{thm}[Basic alternative] \label{res:basic}
	Fix a matrix $A \in \R^{m \times n}$ and a vector $b \in \R^{m}$. Exactly one of the following holds.
		\begin{enumerate}
			\item [P.] There exists $x \in \R^n$ satisfying
			\[
			A x \leq b.
			\]
			\item[D.] There exists $\l \in \R^m$ satisfying
			\begin{align*}
				\l \geq 0, \qquad
				\l^T A = 0, \qquad
				\l^T b < 0.
			\end{align*}
		\end{enumerate}
\end{thm}

In the main text, I show that the primal and dual systems in \cref{res:unified} cannot both be feasible. Here, I show that at least one of the two systems is feasible. To apply \cref{res:basic}, the trick is that strict inequalities can always be scaled up until the gap between the two sides is at least $1$. Formally, the primal system (without the nonnegativity constraint) in \cref{res:unified} is feasible if and only if there exists $(x,t) \in \R^{n + 1}$ satisfying
\begin{align*}		
	t &\geq 1, \\
	A_1 (tx) &= t b_1, \\
	A_2 (tx)&\leq  t b_2, \\
	A_3  (tx)  &\leq t b_3, \\
	\mathbf{1}^T A_3 (tx) + 1 &\leq \mathbf{1}^T t b_3,\\
	A_4 (tx) + \mathbf{1} &\leq  t b_4.
\end{align*}
With the change of variables $u = t x$, we can see that this statement holds if and only if there exists $(u,t) \in \R^{n + 1}$ satisfying:
\[
\begin{bmatrix}
	0 &  -1 \\
	A_1 &  - b_1 \\
	-A_1 & b_1 \\
	A_2 & - b_2 \\
	A_3 & - b_3 \\
	\mathbf{1}^T A_3 & - \mathbf{1}^T b_3 \\
	A_4 & - b_4
\end{bmatrix}
\begin{bmatrix} u \\ t \end{bmatrix}
\leq
\begin{bmatrix}
	-1 \\
	0 \\
	0\\
	0\\
	0 \\
	-1\\
	-\mathbf{1}
\end{bmatrix}.
\]
Denote the matrix on the left by $\tilde{A}$ and the vector on the right by $\tilde{b}$. If this system is infeasible, then by \cref{res:basic} there exists 
\[
\l = ( \l_0, \l_1^+, \l_1^-, \l_2, \hat{\l}_3, \bar{\l}_3, \l_4) \in \R^{1 + m_1+ m_1 + m_2 + m_3 + 1 + m_4}
\]
satisfying
\begin{equation}\label{eq:lambda}
	\l \geq 0, 
	\qquad
	\l^T \tilde{A} = 0, 
	\qquad
	\l^T  \tilde{b}  < 0.
\end{equation}
Let $\l_1 = \l_1^+ - \l_1^-$ and $\l_3 = \hat{\l}_3 + \bar{\l}_3 \mathbf{1}$. Then \eqref{eq:lambda} becomes
\begin{align*}
	\l_i &\geq 0~\text{for}~i = 0,2,3,4, \\
	\sum_{i=1}^{4} \l_i^T A_i &= 0, \\
	\sum_{i=1}^{4} \l_i^T b_i &=  -\l_0, \\
	-\l_0  - \bar{\l}_3 - \l_4^T \mathbf{1} &<0.
\end{align*}
The last inequality implies that: $\l_0 > 0$ or $\bar{\l}_3 > 0$ or $\l_4^T \mathbf{1} > 0$. Condition \eqref{it:S} in \cref{res:unified} follows. Therefore, $(\l_1, \l_2, \l_3, \l_4)$ satisfies the dual system in \cref{res:unified}. 

The nonnegativity constraint can be expressed as an  agumented componentwise inequality with 
\[
A'_2 = \begin{bmatrix} A_2 \\ - I_{n} \end{bmatrix}, 
\qquad
b'_2 = \begin{bmatrix} b_2 \\ 0 \end{bmatrix}.
\]
Apply \cref{res:unified} (without the nonnegativity constraint) to this system, writing the dual vector to the modified componentwise inequality as $\l_2' = (\l_2, \hat{\l}_2)$. Eliminate $\hat{\l}_2$ from the two constraints in which it appears, \eqref{it:M} and \eqref{it:L}, to get componentwise inequality  in place of equality in \eqref{it:L}. 

If we drop relation $j$ from the primal system for all $j$ in some proper subset $J$ of $\{1,2,3, 4\}$, then the corresponding components of $\l$ are dropped from \eqref{eq:lambda}, which is equivalent to setting those components to zero, as specified in the theorem statement. 

\section{Proof of basic version using elimination} \label{sec:proof_basic}

Recall that Gaussian elimination sequentially eliminates variables from a system of linear equalities. Fourier--Motzkin elimination is the anaologus procedure for a system of linear inequalities, though the number of relations increases at each step, making the algorithm computationally inefficient.\footnote{Efficient linear programming algorithms are available for computing solutions of large systems of linear inequalities, but Fourier--Motzkin is sufficient for the proof.}


\begin{lem}[Eliminating a variable] \label{res:FM} Fix a matrix $A \in \R^{m \times n}$ and a vector $b \in \R^{m}$. There exists an integer $p$ and a nonnegative matrix $E \in \R^{p \times m}$ such that (i) the first column of $EA$ is zero, and (ii) the system $Ax \leq b$ is feasible if and only if the system $EA x \leq E b$ is feasible. 
\end{lem}

I first prove \cref{res:basic} from \cref{res:FM}. The primal and dual cannot both be feasible since this would yield the contradiction $0 < 0$. Therefore, it suffices to check that at least one system is feasible. To show this, iteratively apply \cref{res:FM} to the system $A x \leq b$. At each step, the number of unknowns decreases by one; the feasibility of the system is unchanged; and every new inequality is a nonnegative linear combination of inequalities in the original system. After $n$ steps, we obtain an integer $\bar{p}$ and a matrix $\bar{E}$ such that (i) $\bar{E} A $ is the zero matrix, and (ii) the system $A x \leq b$ is feasible if and only if $0 \leq \bar{E} b$.  Thus, if the primal system $Ax \leq b$ is infeasible, then we have $0 \not\leq \bar{E} b$, that is, $(\bar{E} b)_i  < 0$ for some $i$. Let $\l^T$ be the $i$-th row of $\bar{E}$, which is an $m$-vector. We have $\l \geq 0$ (since $\bar{E}$ is nonnegative), $\l^T A = 0$ (since $\bar{E}A$ is the zero matrix), and $\l^T b = (\bar{E}b)_i < 0$, as desired. 

To prove \cref{res:FM}, partition the rows of $A$ according to the sign of $a_{i1}$: 
\[
	I_0 = \{i: a_{i1} = 0\}, \qquad 
	I_+ = \{ i: a_{i1} > 0\}, \qquad 
	I_- = \{i: a_{i1} < 0\}.  \qquad
\]
The vector inequality $Ax \leq b$ can be expressed as
\begin{align}
	& \label{eq:old_1} a_{i2} x_2 + \cdots + a_{i n} x_n \leq b_i, \quad i \in I_0 \\
	& \label{eq:old_2} x_1 \leq  a_{k1}^{-1} (b_k - a_{k2} x_2 - \cdots - a_{k n} x_n), \quad k \in I_+ \\
	&\label{eq:old_3} x_1 \geq  a_{\ell 1}^{-1} (b_\ell- a_{\ell 2} x_2 - \cdots - a_{\ell n} x_n), \quad \ell\in I_-.
\end{align}
For any fixed $x_2, \ldots, x_n$, there exists $x_1$ such that $(x_1, x_2, \ldots, x_n)$ satisfies the $|I_+ \cup I_-| $ inequalities in \eqref{eq:old_2}--\eqref{eq:old_3} if and only if $(x_2, \ldots, x_n)$ satisfies the $|I_+ \times I_- |$ inequalities\footnote{The forward implication is immediate. For the backward implication, observe that any solution $(x_2, \ldots, x_n)$ of \eqref{eq:new_system} satisfies
	\[
	\max_{ \ell \in I_-} a_{\ell 1}^{-1} (b_\ell - a_{\ell 2} x_2 - \cdots - a_{\ell n} x_n) \leq \min_{k \in I_+} a_{k1}^{-1} (b_k - a_{k2} x_2 - \cdots - a_{k n} x_n).
	\]
For any $x_1$ between these bounds, the vector $(x_1, x_2, \ldots, x_n)$ satisfies \eqref{eq:old_2}--\eqref{eq:old_3}.}
\begin{equation} \label{eq:new_system}
a_{\ell 1}^{-1} (b_\ell - a_{\ell  2} x_2 - \cdots - a_{\ell n} x_n) \leq  a_{k1}^{-1} (b_k- a_{k2} x_2 - \cdots - a_{k n} x_n), \quad (k. \ell) \in  I_+ \times I_-.
\end{equation}
Combine \eqref{eq:old_1} with \eqref{eq:new_system} to get a system of $| I_0| + | I_+| |I_-|$ inequalities involving only $x_2, \ldots, x_n$ that is feasible if and only if the system \eqref{eq:old_1}--\eqref{eq:old_3} is feasible. The new system can be written in matrix form as $EA x \leq E b$, where $E$ consists of the rows $e_i^T$ for all $i \in I_0$ and $a_{k1}^{-1} e_k^T -a_{\ell 1}^{-1} e_\ell^T$ for all $(k, \ell) \in I_+ \times I_-$. The first column of $EA$ consists of all zeros (by the definition of $I_0$) and $E$ is nonnegative (by the definitions of $I_+$ and $I_-$). 

\bibliography{lit_theorem_alternative.bib}
\bibliographystyle{ecta}

\end{document}